# The Origins of Lattice Gauge Theory

K.G. Wilson

Smith Laboratory, Department of Physics, The Ohio State University, 174 W. 18th Ave., Columbus, OH 43210

## 1. INTRODUCTION

This talk is an anecdotal account of my role in the origins of lattice gauge theory, prepared for delivery on the thirtieth anniversary of the publication of my article called "Confinement of Quarks" in 1974 [1]. The account is intended to supplement prior books on the history of elementary particle theory in the 1960's and 1970's, especially the book by Andrew Pickering called *Constructing Quarks* [2]. Another reference is a more recent history by Hoddeson *et al.* [3]. The book of Pickering is especially useful because it discusses how a number of physicists developed expertise in one specific aspect of elementary particle physics but then had to adapt to new developments as they occurred. Pickering makes clear that each physicist in the story had to acquire new forms of expertise, while building on the expertise each had already acquired, in order to pursue these developments. But he did not give a full account of the expertise that I developed and used in my contributions to the subject. He provided only a few details on the history of lattice gauge theory, all confined to a single footnote (see [3] for more on the history of lattice gauge theory). This talk fills in some of the gaps left in Pickering's history.

I also describe some blunders of mine, report on a bizarre and humorous incident, and conclude with some concerns. Our knowledge of the true nature of the theory of strong interactions is still limited and uncertain. My main worry is that there might be currently unsuspected vector or scalar colored particles that supplement color gluons and that result in unsuspected additional terms in the QCD Lagrangian for it to fit experiment. I also worry that there is not enough research on approaches to solving QCD that could be complementary to Monte Carlo simulations, such as the lack of any comparable research build-up on light-front QCD. I share the concern of many about how to justify continued funding of lattice gauge theory, and of high-energy physics overall, into the far future: see the end of this talk.

I note that over the past few years I have spent more time researching the history of science than I have on physics. I am particularly indebted to the Director and staff of the Dibner Institute for the History of Science and Technology, at MIT, for the award of a fellowship for the Fall of 2002. The Dibner Institute has a project known as the HRST project that includes an interview with me about my work on renormalization in the 1960's, work that will be touched on later in this talk. On this latter part of my history, a more extensive account is provided in [4]. This talk is informed by my experience with historical research, although it is intended to have the anecdotal flavor that physicists expect in such talks.

This talk is divided into six further sections. The second section is a bow to the present state and future prospects for lattice gauge theory. These prospects seem considerably rosier today than they were when I ceased my own involvement in lattice gauge research around 1985. The third section is about the period in 1973 and 1974 during which I wrote my 1974 article. The fourth section is about the earlier period of my research from 1958 to 1971, beginning with my thesis project suggested by Murray Gell-Mann. In the fifth section I report on blunders of mine after 1970, and also report on a bizarre episode that occurred at that time. In the sixth section I raise some questions for research, including the issue of possible partners for the gluon. A conclusion ends this talk.

## 2. HOMAGE TO LATTICE GAUGE THEORY TODAY

The current knowledge base in lattice gauge theory dwarfs the state of knowledge in 1974, and even the state of knowledge in 1985. The accuracy and reliability of lattice gauge computations is vastly improved thanks in part to improved algorithms, in part to increased computing power, and in part to the increased scale of the research effort underway today. The breadth of topics that have been researched is also greater today than in earlier decades, as one can surely tell by comparing these Proceedings with any similar conference proceedings from twenty or more years ago. But this does not mean that the present state of lattice gauge computations is fully satisfactory. The knowledge and further advances that will likely accumulate over the next thirty years should be just as profound as the advances that have occurred over the last thirty years. Researchers over the next thirty years should have the help of further advances in computing power that, as best as can be foreseen today, are likely to be as profound as the technological advances of the past twenty or more years.

Thirty years can seem like an eternity to anyone who actually lives through thirty years of adulthood. But for the development of physics, thirty years is only a moment. The oldest part of physics and astronomy, namely research on the motions of the Earth, Moon and Sun, dates back over 2700 years (!) to the still valuable data of the Babylonians, recorded on durable tablets. See, *e.g.*, the article on this topic by Gutzwiller [5]. The main change over the 2700-year history is that the accuracy of observation and prediction both improved repeatedly over the whole 2700-year period, although observation dominated prediction after Copernicus. The improvement after Copernicus was aided by profound advances in instrumentation, made possible in part by profound changes in society as a whole as today's advanced economy emerged from the far more limited economy of the time of Copernicus. For example, Galileo's telescope was little more than a spyglass; repeated improvements since then have now resulted in the Hubble Space Telescope and even more powerful ground-based instruments. In the case of research on the Earth-Moon-Sun system, see [5] for more details.

I will not try to imagine what lattice gauge theory would be like 2700 years from now. Could it remain a topic of research for as long as research on the Earth-Moon-Sun system has lasted? If so, what accuracy could conceivably be achieved over this long time period? But it is worth noting that it would have been very difficult for researchers of several centuries ago to imagine the accuracy achieved today in Earth-Moon-Sun studies, or in astronomy and physics more generally, and especially to conceive of the extent and quality of instrumentation that underlies today's accuracy. It would have been equally difficult to predict the economic changes that have occurred over the same period. Surely it is just as difficult to forecast what the circumstances could be for scientific research over two millennia into the future from now.

Nevertheless, I still remember a talk on the future of physics delivered by Richard Feynman at MIT in the early 1960's. He followed Rudolf Peierls, who looked ahead about a century, based on the state of physics about a century earlier, in 1860. According to my possibly faulty memory, Feynman proposed to look forward a thousand years, saying that he was completely safe in his attempt because no one alive to hear his talk could prove him wrong. He said that he could not use history to help him because who was doing physics in 960? All I remember from his predictions was that one possibility was that so many discoveries would occur over the next thousand years that physicists would become bored with making more such discoveries. He discussed other possibilities as well, which were surely more consistent with the fact that astronomers have yet to become bored with research on the solar system after around 2700 years of research.

## 3. FROM ASYMPTOTIC FREEDOM TO LATTICE GAUGE THEORY (1973-4)

The seminal papers on asymptotic freedom by Gross and Wilczek [6] and Politzer [7] were both published in June of 1973, but their content had spread in preprint form earlier. The discovery of asymptotic freedom, made possible by earlier developments on the renormalizability of non-Abelian gauge theories by Veltman and 't Hooft [8], made it immediately clear, to me as well as many others, that the preferred theory of strong interactions was quantum chromodynamics (QCD). The Lagrangian for QCD had already been written down by Gell-Mann and Fritzsch [9].

Unfortunately, I found myself lacking the detailed knowledge and skills required to conduct research using renormalized non-Abelian gauge theories. My research prior to 1973 had not required this knowledge so I had never spent the time necessary to acquire it.

What was I to do, especially as I was eager to jump into this research with as little delay as possible? I realized that from my prior work in statistical mechanics (see Sect. 4) I knew a lot about working with lattice theories, including the construction of high temperature expansions for such theories. I decided I might find it easier to work with a lattice version of QCD than with the existing continuum formulation of this theory. Moreover, this meant I could be doing original research immediately, rather than having to spend weeks or months absorbing other people's research. (I did not learn until much later that a graduate student at UCLA named J. Smit had already been formulating lattice gauge theory, and that Polyakov also worked it out independently of me: see [10].)

Formulating the theory on a lattice turned out to be straightforward, and by sometime in the summer of 1973 I was far enough along to give a talk on lattice gauge theory at Orsay. (I do not remember this talk today, but it is referenced in an article by Balian, Drouffe, and Itzykson [11]. Finding such highly credible and contemporary clues to a scientist's activities is, I learned, a major goal of historical research.) By summer, I knew that:

- The gauge theory lives on a Euclidean four-dimensional lattice.

- The gauge variables are located on nearest neighbor links of the lattice and must be unitary matrices from the gauge group itself, rather than being variables from the infinitesimal Lie algebra of the group similar to the gauge variables of the continuum theory.

- The quark variables are purely anti-commuting objects (see [1]) living at lattice sites.

- There are nearest neighbor couplings of the quark variables with a coefficient denoted $K$, and plaquette-type couplings of the gauge variables with a coefficient of order $1/g^2$.

- There is a separate gauge invariance of the theory for each lattice site.

- The theory with a non-zero lattice spacing gives finite results without any requirement for gauge fixing terms.

However, the concept of confinement was *nowhere in my thinking* when I started my effort to construct lattice gauge theory. I had no idea that confinement would be the central and the most interesting outcome of this research. Even the lack of need for gauge fixing only became apparent after I had worked out the formulation of the lattice theory.

According to my memory of events from thirty years ago, once I wrote down the lattice Lagrangian [1], it was immediately obvious that it would have a strong coupling expansion because the coefficient of the gauge plaquette term was $1/g^2$, rather than $g^2$. Moreover, for quarks, the lattice theory had both a single site term and a nearest neighbor term, and the coefficient $K$ of the nearest neighbor term could also be made artificially small if I chose to do so. (For my notation, see Eq. (3.12) of [1].) But when I started to study the strong coupling expansion, obtained by assuming $g$ is large and $K$ is small, I ran into a barrier. I could carry out the mathematical manipulations required to produce the strong coupling expansion with ease. But for many months I could not turn the mathematics into a physical interpretation that I could understand. My problem, as best I can mentally reconstruct my struggles of so long ago, was that I did not spend enough time studying the very simple limit in which all purely space-like terms with small coefficients are dropped altogether from the Lagrangian. This simplification leaves only gauge field plaquettes in one space and the time direction and nearest neighbor quark couplings in the time-like direction. In this simple limit there are a number of crucial results, all resulting from analysis of the transfer matrix formalism linking the lattice Lagrangian to the system's Hamiltonian (see [12] for this formalism), namely:

- All non-gauge-invariant states, such as single quark states, have infinite energy.

- The elementary particles of the theory are quarks that live at single lattice sites and "string bits" that live on space-like nearest neighbor sites.

• The masses of gauge-invariant composite multi-particle states are sums of the masses of each constituent, with the mass of a quark constituent being $-\ln(K)$ and the mass of a constituent string bit being $-2\ln(g)$. For $K$ and $g$ small enough, both these masses are large, indeed much larger than the inverse of the lattice spacing, which is taken to be unity in these formulae. This rule includes the appearance of strings built of string bits whose total mass is proportional to the length of the string.

These simple results are all easily verified from the form of the transfer matrix when $K$ and $g$ are small and when purely space-like terms in the action are dropped. But rather than focus on this simple limit, I spent months trying to make sense of the vast number and the complexity of the diagrams that arise when purely space-like terms are taken into account rather than neglected. I would stare at the rules for constructing the strong coupling expansion (large $g$, small $K$) and wonder what examples I should actually compute in order to improve my understanding of it, but be unable to identify a specific computation that was likely to help me. Then I would spend time studying various papers on string theory, including one by Susskind, and at least one of the papers of Kogut and Susskind, and make only a little progress in extracting useful insights from their work. But the situation did eventually become clarified enough so that I was able to write the article now being celebrated. This article was received by the Physical Review in June of 1974, almost a year after my talk at Orsay.

The resulting article says nothing about the struggles I went through to produce it, and indeed covers rather limited ground altogether. The detailed rules for small $K$ and $g$ stated above are absent. The basic formalism is present, along with the concept of gauge loops and the area law for large gauge loops implying quark confinement. The lack of any need for gauge fixing was noted. Connections to string theory are noted, including the appearance of diagrams involving sums over surfaces: but the article questions whether it will be easy to link sums over discrete surfaces on the lattice to sums over continuum surfaces in string theory. A few other topics are dealt with by referring to other papers, mostly not my own.

From its inception, I was not alone in having the skills needed to conduct research on lattice gauge theory. The lattice gauge theory was a discovery waiting to happen, once asymptotic freedom was established. Franz Wegner published an earlier paper on an Ising-like theory with plaquette couplings that has local gauge invariance [13]. My 1974 article has references to papers then about to be published by Balian, Drouffe, and Itzykson [11] and Kogut and Susskind [14]. If I had not completed and published my work in a timely fashion, then it seems likely that Smit, Polyakov, or both [10] would have produced publications that would have launched the subject.

In any case, while I have focused on the history of my own article, a true and more extensive history of the origins of lattice gauge theory is a history of a number of individuals who contributed to the initial set of articles on lattice gauge theory in 1974 and 1975. Some of those individuals are mentioned above. Others contributed to the history of topics that provided the necessary background for these articles, from the pioneering work on the renormalizability of non-Abelian gauge theories to the statistical mechanical background that I drew on. See, *e.g.*, Refs. [2, 3].

This discussion leaves a question to answer: how and when did I acquire experience with high temperature expansions in statistical mechanics, which prepared me to think about lattice theories and the similar strong coupling expansion for lattice gauge theory? It happens that two of the leading researchers on critical phenomena in statistical mechanics of the 1960's were in the Chemistry Department at Cornell, namely Michael Fisher and Ben Widom, and they introduced me to the challenges of this topic. I learned from listening to them that the Ising model could serve as a "theorists' laboratory" for the investigation of the same kind of phenomena that lead to renormalizability challenges in quantum field theory. The phenomena in question include the phenomenon of anomalous dimensions that had yet to be thought about in field theory. Such anomalous dimensions had already been found to occur in the exactly soluble Thirring model [4], but there had been no follow-up of this curious result. This leads me to the next topic of this talk: my earlier work of the 1960's, including work on a form of the renormalization group that seems less likely to have been developed by someone else had I not pursued it.

## 4. 1958-1971: A THESIS PROJECT AND ITS AFTERMATH

The history of my role in the origins of the lattice gauge theory was a short one, confined to a single year. But one cannot understand how I came to play the role that I did until one understands my earlier history as a physicist over a far longer period of time: from 1958 to 1971. Remarkably, I wandered off the beaten track of physics for much of this very long period, yet at the end of it, I was able to produce results that would surely have taken considerably longer to be generated by someone else in my absence. It is also a very haphazard history; it has no organized plan to it, although in retrospect I can make it sound more logical than it actually was. The outcomes of this history could not have been anticipated in advance. This section is, in part, an abbreviated version of a more extensive historical discussion that I provided in 1983 [4]. See also a recent review by Fisher [15] and the materials and interviews collected by the Dibner Institute's HRST project [16].

In 1958, I was given a project to work on for a thesis by Murray Gell-Mann. I pursued this topic in a direction that was different from his primary interest. Like many second-rate graduate students, I pursued ideas from my thesis topic for over fifteen years before disengaging from it. This topic began as an application of the Low equation in the one-meson approximation to $K$-$p$ scattering. In the one-meson approximation, the Low equation was a one-dimensional integral equation with two explicit parameters: a meson mass, and a meson-nucleon coupling constant.

Murray's goal was to use the equation to help make sense of the phenomenology of $K$-$p$ scattering. But I became fascinated with the high-energy behavior of solutions to the Low equation, despite its being a reasonable approximation for physics, if at all, only for low energies. In this way, I embarked on ten years of research that was seemingly irrelevant to physics; but it led me to a very distinctive way of thinking that underlies my work on the renormalization group published in the early 1970's, and also prepared me to take an interest in the Ising model and critical phenomena [4].

A crucial aid to my research was the use of digital computers. Jon Mathews, a faculty member at Cal Tech when I was a graduate student, introduced me to Cal Tech's 1950's computer—a Burroughs Datatron that was programmed in assembly language. From analytical studies, I learned that the scattering amplitude f$(E)$ that solves the Low equation has an expansion in powers of $g^2 \ln(E)$ for large energy $E$. Using the computer, I learned that the coefficients of this expansion are integers $a(n)$ multiplying the $n$th power of the logarithm. Then I was able, to my surprise, to find an analytic form for these coefficients valid for all $n$. This work was in my thesis and later written up as a preprint which I apparently never published.

I also found that the renormalization group concepts of Gell-Mann and Low [17] were applicable to the high-energy behavior of the Low equation, which only strengthened my interest in this topic. I wrote a thesis with my results up to 1961, and, because Gell-Mann was on leave in Europe, Feynman read my thesis. I gave a talk on it at the theory seminar led that year by Feynman. As one would expect, that seminar generated a Feynman anecdote. After I finished my talk, someone in the audience asked: "Your talk was interesting, but what good is it?" Feynman answered, saying "Don't look a gift horse in the mouth!" Feynman's answer was prophetic, but it was not until 1971 that any real payoff came from further research building on my thesis results.

Two years after turning in my thesis, in 1963, I decided to focus my research on the high-energy behavior of quantum field theory. I knew that a theory of strong interactions would be a theory not soluble by perturbation theory because the interactions involved were too strong. I also knew that no imaginable theory of the time made enough sense to be correct. (Quarks had yet to be proposed, let alone taken seriously as constituents of hadrons. For the difficult history of the quark hypothesis and of early advocates of quarks between 1964 and 1973, see [2].) But my main motivation in entering physics—a decision I made in high school—was to have interesting and productive mathematical problems to solve. By 1963, I concluded that the most interesting and useful mathematical problem for me to work on was the question of high-energy, large-momentum transfer behavior in quantum field theory, which linked up to issues of renormalizability and the renormalization group ideas of Gell-Mann and Low. With this decision, I became largely isolated from the mainstream of high-energy physics of the time, working instead largely in a world of my own [2, 4].

My decision about how to focus my research was an example of what Gerald Holton, a historian of physics, calls a "thematic presupposition." A thematic presupposition is a supposition or hunch about what some area of physics is like, or about what kind of research will pay off, made before there is adequate evidence to justify the hunch. Any scientist entering a new line of research has to rely on such suppositions or hunches, as Holton makes clear: see [18] and [19]. (In [18], Holton has a fascinating discussion of the failure of Kepler to persuade Galileo to become enthusiastic about Kepler's ellipses, despite repeated attempts by Kepler to attract Galileo's attention. The early advocates of quarks had an equally difficult time getting respect for their concept, as Pickering documents in chapter 4 of [2].)

I drew on a number of "theorist's laboratories" for my research, including the Low equation and the more complex static fixed source theory of mesons interacting with a fixed, point-like nucleon that underlay the Low equation. Another laboratory was the high-energy behavior of full field theories solved in perturbation theory. Around 1965, I added the Ising model to this list of laboratories for my research, as I learned that renormalization group ideas were as relevant for the Ising model as for quantum field theory, in particular through an approach to the Ising model developed by Leo Kadanoff (see [4]).

I mastered, at least partially, as many approximation methods as I could find that might be useful, one being high temperature expansions for the Ising model. These expansions were used in the 1960's to extract critical exponents [15], which I learned were related to the anomalous dimensions of quantum field operators in quantum field theory. Indeed, the Ising model could be interpreted as a quantum field theory defined on a discrete lattice rather than in the continuum, with the help of the transfer matrix formalism that had already been developed as a route to solving the Ising model [20].

I struggled to come up with something truly useful from my research, but at the same time was happy that I was making progress without getting to useful results too quickly. Getting results too quickly would mean that the problems I was working on were not difficult enough to be as challenging as I wanted them to be.

I will outline one of my most helpful "experiments" conducted in a theorist's laboratory. Inspired by my thesis work, I was studying the fixed source theory, in which an interaction is remarkably simple to write down. I used:

$$g\,\boldsymbol{\phi}(0)\bullet\boldsymbol{\tau}$$

where $g$ is the coupling constant, $\boldsymbol{\phi}$ is a scalar and isovector meson field, and the $\boldsymbol{\tau}$ matrices are the isospin matrices of a nucleon fixed at the origin in space [21]. The Hamiltonian also included a relativistic free field Hamiltonian for a meson field of mass $m$.

In order to understand the nature of renormalization for this theory, I decided to study a butchered form of this theory that opened up a new way of approximating its solution. The butchered form I called a "momentum slice" theory, in which mesons could exist only in discrete momentum intervals rather than at any momentum. The allowed momentum ranges [21] could be, *e.g.*,

$$1 < |\boldsymbol{k}| < 2;\ 1000 < |\boldsymbol{k}| < 2000;$$
$$1{,}000{,}000 < |\boldsymbol{k}| < 2{,}000{,}000, \ldots$$

The free field operator $\phi(\boldsymbol{x})$ in the butchered theory was an integral of a linear combination of creation and destruction operators:

$$\phi(x) = \int_k \left(a_k e^{ik\cdot x} + a_k^+ e^{-ik\cdot x}\right) \tag{1}$$

but in the integration, $\boldsymbol{k}$ is a vector whose magnitude is confined to the discrete momentum ranges defined above rather than varying continuously from 0 to infinity.

There could be a finite set of these ranges, thereby providing a cutoff form of the theory with a maximum momentum (call it $\Lambda$) for the theory, or there could be an infinite number of these momentum slices, corresponding to the case of no cutoff. The distinct momentum scales had distinct energy scales as well. Assume the meson mass m is of order 1. Then the meson energy in each slice is of order of the momentum in each slice, with the meson energy in the first slice of order 1, in the second slice of order 1000, and in the third slice of order 1,000,000.

Whenever a quantum mechanical Hamiltonian involves multiple and well separated energy scales, there is a standard rule for how one solves it: one starts by examining the part of the Hamiltonian involving the largest energy scale, ignoring all smaller energies. Then one brings in the terms involving the second largest scale, treated as a perturbation relative to the largest scale. This is followed by bringing in the terms of order the third largest energy scale, treated as a perturbation with respect to both the largest and second-largest scales. The details for the momentum slice model are discussed in [21].

The result of carrying out this orderly procedure for solving the Hamiltonian was a renormalization group transformation in a novel and unexpected form. To a reasonable first approximation, the result of solving the Hamiltonian on the largest energy scale and then bringing in the next largest scale amounted to the generation of a self energy on the largest scale for the nucleon, then dropping the largest energy scale terms from the Hamiltonian, but with the replacement of $g$ by a renormalized constant $g_1$ that is related to $g$ through a multiplicative factor that depends on $g$ (to be precise, the multiplicative factor is obtained from the ground state expectation value of $\tau$ computed using the ground state of the unperturbed largest scale term in the original Hamiltonian: see [21].) That is, the original Hamiltonian with a top momentum of $\Lambda$ is replaced by the same Hamiltonian with a new coupling constant $g_1$, a lowered cutoff of $0.001\Lambda$ and a self-energy term of order $\Lambda$ for the nucleon. The connection of $g_1$ to $g$ can be written in the form

$$g_1 = g w(g) \tag{2}$$

where $w(g)$ is a matrix element [21], the detailed form of which need not concern us.

Continuing the procedure brings about further reductions in the cutoff, along with further changes in the coupling constant. That is, solving the largest term in the Hamiltonian with $g_1$ resulted in a Hamiltonian with the second-to-largest energy scale also removed: the new cutoff is $10^{-6}\Lambda$ with a second new coupling $g_2$ given by:

$$g_2 = g_1 w(g_1) \tag{3}$$

The result of this analysis, so far, is a Gell-Mann-Low type renormalization group, although with a recursion equation giving changes in $g$ for discrete reductions in the cutoff rather than a differential equation for $g$. See [21] for a more extended discussion, including detailed equations for Hamiltonians and relevant matrix elements.

But the most startling result from the momentum slice Hamiltonian came when I solved the theory beyond a first approximation. The result that only the coupling constant changed was valid only if I treated lower energy scales to first order in perturbation theory. Given that the ratio of energy scales was 1000 in my model, it was a reasonable first approximation to stop at first order. But for increased accuracy, one should consider second and higher order terms, *e.g.*, treating terms with an energy scale of $0.001\Lambda$ to second or higher order relative to the energy scale $\Lambda$. Once I took higher order terms into account, I still obtained a sequence of Hamiltonians $H_1$, $H_2$,..., with progressively more energy scales removed. But now the form of the Hamiltonians was more complex, with a considerable number of terms containing new coefficients rather than just a change in g occurring. Moreover, if the computation was carried out to all orders in the ratio of adjacent energy scales, then all the successive Hamiltonians in the sequence contained an infinite set of distinct terms with distinct coefficients. What one had in this case was a renormalization group transformation, to be denoted by $T$, that replaced one Hamiltonian by another, beginning with

$$H_1 = T(H), \tag{4}$$

and

$$H_2 = T(H_1). \tag{5}$$

The transformation $T$ takes the form of a non-linear transformation depending on all the coupling coefficients that appear in the successive Hamiltonians, with the number of coefficients becoming infinite if the transformation is computed exactly.

In the case that the ratio of successive energy scales is 1000, the extra coefficients have only a very minor role in the transformation, with the simple recursion formula for $g$ being the dominant component in the transformation. But if one wants to avoid the artificiality of the momentum slice model, then one has to define momentum slices with no gaps, *e.g.*,

$1 < |\mathbf{k}| < 2;\ 2 < |\mathbf{k}| < 4;\ 4 < |\mathbf{k}| < 8,\ \dots$

In this case the ratio of energy scales is only 2 instead of 1000, and it is no longer clear that one can use a perturbative treatment at all, let alone drop all but the lowest order terms. In this case, if anything at all could be done through perturbation theory, it would clearly involve the use of high orders of perturbation theory, with the result that the transformation $T$ would necessarily generate very complex Hamiltonians with very many coefficients to determine.

In retrospect, the discovery of this new form of the renormalization group, with complex Hamiltonians being transformed into other complex Hamiltonians, was a major advance. It provided me with a way to formulate the renormalization group that could be applied to theories in strong coupling—there was no requirement that $g$ be small, unlike the Gell-Mann–Low formulation which was completely tied to perturbation theory in $g$. But at the time there was nothing I could do with this idea. It potentially established a new world of research, but one that was so ugly that only its mother—namely me—could love it. It is not surprising that I could not do anything with this idea very quickly, and I focused for a while on other aspects of my research.

Somewhat to my surprise, in the early 1970's I was able to extract two very practical payoffs from the new world of renormalization group transformations involving very many couplings rather than just one. I will skip the remainder of the history leading up to these payoffs: see [4,15,16] for some remaining details of this history.

The first payoff, obtained in 1971 and 1972, was the result that is now the dominant factor in my overall reputation as a physicist. The result came in two stages, the second being considerably more useful than the first. The first stage result was a very crude approximation to a renormalization group transformation that had intermediate complexity. The transformation took the form of a one-dimensional integral equation for a very restricted class of Hamiltonians for simple models of critical phenomena in statistical mechanics. But for the benefit of the audience, the transformation can equally be understood to be a transformation for a restricted class of actions for a scalar field $\phi$ in an arbitrary space-time dimension $d$. The restricted class of actions includes a free field term and a completely local interaction term that involves all even powers of $\phi$, not just $\phi^2$ and $\phi^4$. Odd powers could be added, too, as needed. If one denotes this sum of powers as a function $u$ of a scalar variable $\phi$, then the transformation has the generic form

$$u_1(\phi) = T(\phi, u, d) \qquad (5)$$

where $T$ is defined in terms of an integral involving the function $u$ along with other manipulations. See [22] for the details. This transformation turned out to be straightforward to program; moreover, I had access to an empty PDP10 that was installed at Newman Lab and brought to working order just a few weeks before I arrived at this intriguing transformation and wanted to learn more about it.

The second stage of the payoff came when Michael Fisher and I jointly recognized that the one-dimensional integral equation could be solved analytically in a perturbation expansion in $d$ about $d = 4$. In the perturbation expansion, the function $u$ had an unperturbed $\phi^2$ term, and the leading perturbative term was a $\phi^4$ term with a coefficient of order $(4 - d)$. See [23] for details. But this led to the realization that critical exponents could be computed in an expansion in powers of $(4 - d)$ using exact Feynman diagrams computed for non-integral dimension $d$ [24], bypassing the rough approximations that underlie Eq. (5). Many theorists already knew how to compute Feynman diagrams, and the changes needed to bring in a non-integer dimension were quite minor [24]. Moreover, the expansion in $(4 - d)$ could be applied to all kinds of statistical mechanics computations, such as equations of state and various kinds of scattering phenomena [15]. There was already a powerful community of researchers on critical phenomena that had identified a broad variety of phenomena worth studying. The result was an explosion of productive research that built up over the subsequent decade, in striking contrast to the slower build-up of research on lattice gauge theory, which was limited in the years after 1974 because of limitations in computing power and in algorithms for using what computing power was available. See [4] for a variety of further references on the explosion of research in critical phenomena following the second stage of the breakthrough achieved by Fisher and myself [23,24].

In retrospect, I believe that the expansion in $(4 - d)$ for critical phenomena was also a discovery waiting to happen, although it is not clear how or when it would have happened if my research had not occurred. What is important is that at the time there were other physicists familiar with both critical phenomena and quantum field theory, from Migdal and Polyakov to Jona-Lasinio and DiCastro (the latter two were already exploring the applicability of the Gell-Mann-Low renormalization group to critical phenomena: see [4]). The concept of using non-integer dimensions was in the air, too, because of its use for dimensional regularization of gauge field theories: see [4].

But there is also now a broader and very elegant renormalization group framework based on the concept of renormalization group transformations acting on an infinite dimensional space of Hamiltonians (of the kind that appear in Boltzmann factors in statistical mechanics). The elegance begins with the concept of a renormalization group fixed point: a Hamiltonian $H^*$ that satisfies

$$H^* = T(H^*). \tag{6}$$

The formalism includes the concept of relevant and irrelevant operators, in particular as developed by Franz Wegner. See [15] for a recent review that puts this formalism into perspective. I believe that even this formalism would have emerged eventually if my work had never occurred, because of the need to define and learn about irrelevant operators near fixed points, and because of the availability of researchers capable of developing this concept. But it could have taken a number of years for the expansion in $(4 - d)$ and the fixed-point formalism to emerge in the absence of my research. See [15] for a recent review of the impacts of the renormalization group ideas on research in critical phenomena; see also the interviews conducted by the HRST project staff of the Dibner Institute [16].

The other payoff from my work in the 1960's was, if anything, even more of a surprise. It was an approximate numerical solution of a model of an isolated magnetic impurity in a metal, called the Kondo model or Kondo problem. This model was very similar in structure to the fixed source model that was the starting point for the momentum slice analysis discussed above; the main difference was that a free electron (fermion) field was coupled to an impurity treated as a fixed source. The electron energy was a simplified form of the electron energy near the Fermi surface of a metal: see [25]. The numerical approach is based on a momentum slice formulation of the model, but with no gaps between slices, and uses a strictly numerical approximation scheme that avoids any use of the perturbative expansion outlined above. The scheme required the use of a supercomputer of the time, namely a CDC 6600 located at Berkeley that I accessed through a remote batch terminal. For details on the computation and its results, see [25]. I expected the key results to be accurate to around a percent, according to internal consistency arguments. Astonishingly, the model turned out later to have an exact analytic solution, which agreed with my numerical result to within its estimated errors, but just barely. For a recent review of work on the Kondo problem that provides a modern perspective on my work among a number of other topics, see [26]. My work on the Kondo problem is the work that seems least likely to have been produced by someone else if I had not done it. To my knowledge, no one else was thinking about the momentum slice approach to the renormalization group, let alone developing the skills needed to mount a large-scale computation based on it. It was amazing that the computing power of a CDC 6600 was enough to produce reasonably accurate numerical results. Moreover, there is a simple aspect to this work that I never published (which I now regret) and seems not to have been discovered and reported by anyone else. Namely, the Kondo calculation becomes simple enough to carry out on a pocket calculator if one makes a momentum slice model of the Kondo problem with large separations between slices and simplifies each slice. My work on the Kondo problem would surely have been easier for other researchers to understand and build on if I had also published the much simpler form of the renormalization group transformation that results when one butchers the free electron field of the Kondo Hamiltonian to allow only well separated momentum slices in the way that was done for a scalar field in [21].

I have a closing comment. I reviewed my history between 1958 and 1971 in part because it provides background for my work on lattice gauge theory. But I also reviewed it because someday it may prove useful to apply the momentum slice strategy to some other seemingly intractable many-body problems.

## 5. BLUNDERS AND A BIZARRE EPISODE

In the early 1970's, I committed several blunders that deserve a brief mention. The blunders all occurred in the same article [27]: a 1971 article about the possibility of applying the renormalization group to strong interactions, published before the discovery of asymptotic freedom. My first blunder was not recognizing the theoretical possibility of asymptotic freedom. In my 1971 article, my intent was to identify all the distinct alternatives for the behavior of the Gell-Mann–Low function $\beta(g)$, which is negative for small $g$ in the case of asymptotic freedom. But I ignored this possibility. The only examples I knew of such beta functions were positive at small coupling; it never occurred to me that gauge theories could have negative beta functions for small $g$. Fortunately, this blunder did not delay the discovery of asymptotic freedom, to my knowledge. The articles of Gross and Wilczek [6] and Politzer [7] soon established that asymptotic freedom was possible, and 't Hooft had found a negative beta function for a non-Abelian gauge theory even earlier [2].

The second blunder concerns the possibility of limit cycles, discussed in Sect. III.H of [27]. A limit cycle is an alternative to a fixed point. In the case of a discrete renormalization group transformation, such as that of Eq. (6), a limit cycle occurs whenever a specific input Hamiltonian $H^{*}$ is reproduced only after several iterations of the transformation $T$, such as three or four iterations, rather than after a single iteration as in Eq. (6). In the article, I discussed the possibility of limit cycles for the case of "at least two couplings", meaning that the renormalization group has at least two coupled differential equations: see [27]. But it turns out that a limit cycle can occur even if there is only one coupling constant $g$ in the renormalization group, as long as this coupling can range all the way from $-\infty$ to $+\infty$. Then all that is required for a limit cycle is that the renormalization group $\beta$ function $\beta(g)$ is never zero, *i.e.*, always positive or always negative over the whole range of $g$. This possibility will be addressed further in the next section, where I discuss a recent and very novel suggestion that QCD may have a renormalization group limit cycle in the *infrared* limit for the nuclear three-body sector, but not for the physical values of the up and down quark masses. Instead, these masses would have to be adjusted to place the deuteron *exactly at threshhold for binding*, and the di-neutron also [28].

The final blunder was a claim that scalar elementary particles were unlikely to occur in elementary particle physics at currently measurable energies unless they were associated with some kind of broken symmetry [23]. The claim was that, otherwise, their masses were likely to be far higher than could be detected. The claim was that it would be unnatural for such particles to have masses small enough to be detectable soon. But this claim makes no sense when one becomes familiar with the history of physics. There have been a number of cases where numbers arose that were unexpectedly small or large. An early example was the very large distance to the nearest star as compared to the distance to the Sun, as needed by Copernicus, because otherwise the nearest stars would have exhibited measurable parallax as the Earth moved around the Sun. Within elementary particle physics, one has unexpectedly large ratios of masses, such as the large ratio of the muon mass to the electron mass. There is also the very small value of the weak coupling constant. In the time since my paper was written, another set of unexpectedly small masses was discovered: the neutrino masses. There is also the riddle of dark energy in cosmology, with its implication of possibly an extremely small value for the cosmological constant in Einstein's theory of general relativity.

This blunder was potentially more serious, if it caused any subsequent researchers to dismiss possibilities for very large or very small values for parameters that now must be taken seriously. But I want to point out here that there is a related lesson from history that, if recognized in the 1960's, might have shortened the struggles of the advocates of quarks to win respect for their now accepted idea. The lesson from history is that sometimes there is a need to consider seriously a seemingly unlikely possibility. The case of Copernicus has been mentioned. The concept that the Earth goes around the Sun was easily dismissed at the time because of a claim that the stars should not be as far away from the Earth as was needed, namely much farther away than the Sun. In the case of stars not much farther than the Sun, their location in the sky would make easily observable changes as the Earth moves around the Sun. But the reality was that observers in Copernicus' day had no evidence allowing them to actually determine the distance of the nearest star and prove that it was not much farther away than the Sun. There was a second problem with Copernicus' concept that made it even less acceptable. The belief at the time was that bodies in motion would come to a stop in the absence of any force on them. Thus it was assumed that if the Earth was in motion, and a person jumped off the ground, that person would come to a halt while the Earth kept moving, leaving the person behind. But no one who jumps off the ground is left behind in practice. It was not until a century later that precursor forms of Newton's first law of motion became available and helped to remove this concern. (A reference for this history from Copernicus on is [29].)

The concept of quarks, especially as formulated by George Zweig in 1964, suffered similarly, but from three obvious flaws. One was that, as initially proposed by Zweig, it violated the connection between spin and statistics. The second was that it required fractional charges, smaller than the electron charge previously thought to define the smallest possible unit of charge. The third flaw was that it proposed low-mass constituents that should have been easily visible in the final states of scattering experiments, yet none were seen [2].

The violation of the connection between spin and statistics of the original proposal was and remains unacceptable. But within a couple of years, there were proposals to triple the number of quarks in a way that also allowed quark charges to be integers, reducing the number of flaws from three to one [2]. The proposed increase from 3 to 9 in the number of quarks was hard to take seriously until there were data to support it, but the concept could not be ruled out. The final presumed "flaw" turned out to be based on a claim that constituents had to be capable of existing in isolation that is valid only for weakly coupled quantum field theories, which are the only form of quantum field theory that theorists knew anything about at the time. But the lack of knowledge about what was possible in a strongly coupled field theory did not stop many field theorists of the 1960's from being very negative about the quark concept, and I was among those with little enthusiasm for it.

Even today there are areas of physics where we know astonishingly little and where some seemingly preposterous proposals could turn out to be true. To help enable such proposals to be treated skeptically, yet still with respect, I have a suggestion for discussion. My suggestion is that the physics community agree to establish a special category of proposals that seem widely unacceptable but cannot and should not be ruled out. I suggest that this category be called "Pauli proposals" in honor of Pauli's dismissal of an article sent to him because "it was not even wrong!" I suggest that a list of past "Pauli proposals" that ultimately proved successful be compiled and published to help all physicists understand how often such "Pauli proposals" have overcome initial skepticism. Notable past examples include the Copernican theory, the theory of nineteenth century geologists and biologists that the Sun had been shining for hundreds of millions of years, and the quark hypothesis of Zweig. Some nominees for Pauli proposals for here and now will be discussed in the next section.

In support of using Pauli's name for this list of proposals, I have a Pauli anecdote. Pauli gave a seminar at Cal Tech in the late 1950s, when I was there, in which he talked about his perspective on a unified field theory proposed at the time by Heisenberg. He opened the seminar by saying, "I can paint like Titian." He turned to the blackboard and drew a huge rectangle on it. Then he said, "All that is left is to fill in the details."

But his real complaint was that the details Heisenberg did supply were pedestrian. His theory was not crazy enough to be interesting.

To close this section, I turn to a bizarre and humorous incident that occurred in the early 1970's. By then I had published a very daunting article [30] proving that the full transformation T of the momentum slice version of the scalar fixed source theory [see Eqs. (4) and (5)] did not lead to significantly different results than the much simpler leading order approximations of Eqs. (2) and (3). My worry was that through many iterations of the full transformation $T$, the resulting Hamiltonians $H_n$, for $n = 1,2,3,4,...$, would drift steadily away as $n$ increased from the much simpler Hamiltonians of the leading order approximation. With the help of a long sequence of upper bounds on higher order terms, I concluded that no such drift could occur if the ratio of energies in successive momentum slices was small enough. After the publication of this article in the Physical Review [30], I got a phone call from someone who claimed that he had read and admired this article. He also had some thoughts that he wanted to share with me, and offered to come to Ithaca for a private meeting. I was taken aback that anyone would actually read the article in question, and I agreed to a meeting that took place on a Saturday morning in Ithaca, with me staying at the door to Newman Laboratory to let in my visitor. The essence of our conversation was that my visitor had a view of how the world worked with which he was sure I would be sympathetic. His view was that the world we live in is actually a computer simulation conducted in a higher-level world invisible to us. Moreover, he was especially proud of his explanation for quantum mechanical uncertainty. This uncertainty is due to *bugs in the computer simulation program.*

The reason he thought I would be interested in his thinking is that he realized that the higher order world would also be a computer simulation in an even higher-level world. He expected that these higher order worlds exist in a long progression, just like the iterations of the renormalization group that characterized my work. I kept listening. But then he asserted that eventually one comes to a computer simulation run by a Supreme Being. At that point I managed to terminate the conversation. My visitor left, and I have not heard from him again. I try to be open-minded, but I am not prepared to follow up on his proposal, not even as a candidate for the list of Pauli proposals.

## 6. CONCERNS FOR THE FUTURE OF LATTICE GAUGE THEORY

I will now return to the topic of lattice gauge theory, and discuss some concerns that I have about it for the present and the future. My first concern, mentioned in the Introduction, is that the currently accepted QCD Lagrangian may not be the correct one, as discussed below. My second concern is to point out the unexpected possibility that QCD has a limit cycle in the three-nucleon sector, and discuss what this implies for lattice gauge theory. My third concern is to suggest that there needs to be more attention to research on light-front QCD as a complement to research on lattice gauge theory. Finally, I discuss the expense of lattice gauge theory, and indeed of high-energy physics as a whole, and the issue of justifying this expense in an increasingly confusing world.

I have had a concern about the correctness of the QCD Lagrangian from the inception of lattice gauge theory. In my view, a major purpose of developing and trying to solve lattice gauge theory is in order to determine whether the QCD Lagrangian truly *does* account for strong interactions, and if so, to what *accuracy* does it explain experimental results? At the beginning of this article, I saluted the progress that has occurred on lattice gauge theory, which includes computations that seem to agree with experiment to a few percent, as is discussed in other talks in these Proceedings. But how large are the computational errors in the numbers reported due to limited size effects, and especially due to the approximations used to handle quark vacuum polarization effects? I suspect that it is still too soon to be able to provide reliable upper limits for these errors.

However, one serious shortcoming of my own writings about lattice gauge theory has been that I have never expressed my skepticism about the correctness of the QCD Lagrangian in a useful form. I have never suggested what realistic alternatives to the QCD Lagrangian might look like and should be tested for. By now, any concrete proposal for an alternative to QCD would constitute a Pauli proposal, because there is no evidence (to my knowledge) that requires consideration of the alternatives I will discuss. But when one looks at the structure of the known families of elementary particles: the quark family, the two lepton families, and even the vector gauge boson multiplet of the photon, *W* and *Z* particles, one has to ask a question. Does the gluon octet have unsuspected partners? I remind readers that the neutron and the muon, partners for the proton and electron, respectively, were entirely unsuspected at the time they were discovered. Thus there is ample precedent for more partners for the gluon to exist even if they were not anticipated in any theorist's specific model.

There are already serious suggestions for possible partners for the gluon that are discussed in the reports of the Particle Data Group, such as supersymmetric partners, which would be fermions [31], or a separate suggestion for the existence of axigluons (pseudovector gluons) [31]. These suggestions come from specific models that extend the Standard Model, and certainly research on these possibilities should continue. But my view is that, during the next decades, lattice gauge computations should become increasingly accurate and reliable, and as they do, discrepancies with experiment could arise that cannot be explained by either experimental or computational errors. I have asked myself how I would try to explain such discrepancies if they should occur and not be explainable by any of the presently proposed extensions of the Standard Model. It is hard to give an answer without knowing where these discrepancies arise and what their magnitudes are, but I have a possibility to propose even in the absence of such information. The possibility I suggest is that the gluon has heavier partners that are either of the same type (uncharged vector particles that are also color octets) or are scalar rather than vector particles. Unfortunately, I cannot make any guesses as to the masses or the strength of the color couplings of these hypothetical particles, which means that searches for them cannot rule them out altogether but might set various kinds of bounds on their masses and coupling constants. I do not have the time or assistance to do detailed computations of how such particles might change the predictions of normal QCD, and without such information I think a list of Pauli proposals might be the right place for my suggestion. However, if some other physicists do carry out and publish such computations, then they could deserve and receive, serious consideration by experimentalists. Any such physicists should get primary credit if the gluon partners for which they do computations are subsequently discovered.

The next topic is research by nuclear physicists on a possible limit cycle in the nuclear three-body problem. Curiously, the basic theoretical phenomenon of this limit cycle was discovered around 1970 (over thirty years ago) by Efimov [32]. The initial discovery of Efimov is that a non-relativistic system of three equal mass Bose particles with delta function two-body potentials and a suitable cutoff (the theory is not soluble without a cutoff) has an infinite set of discrete bound states with a point of accumulation at zero energy. This comes with the proviso that the two-body system has a bound state exactly at threshhold. Moreover, as the two-body system has bound states come closer to zero energy, the ratio of successive three-body bound state energies approaches a fixed limit of about the reciprocal of 515.03.

The connection to the renormalization group and the possibility of a limit cycle was not recognized until much more recently, starting with a paper by Bedaque, Hammer, and Van Kolck [33] in which a three-body delta function term with a coefficient h was added to the Hamiltonian. They showed that if one tried to renormalize the Hamiltonian with both two-body and three-body delta functions terms, to obtain fixed renormalized eigenenergies while a momentum cutoff was increased to infinity, then the two-body coupling approached a constant. But the three-body coupling $h$ went through an unending limit cycle, cycling from $-\infty$ to $+\infty$ and then jumping to $-\infty$ to repeat the cycle. But the Bedaque *et al.* article [33] does not explicitly recognize that they were seeing a renormalization *group* phenomenon, namely a limit cycle of the kind that I discussed in my 1971 paper, although with only one coupling constant participating in the cycle.

To my surprise, a related recent discovery is that a colleague of mine named Stan Glazek, who spent a year at The Ohio State University, and I had written a paper in the early 1990's that included, in an appendix, a much simpler model Hamiltonian that is exactly soluble and exhibits a very similar renormalization group limit cycle [34]. But our paper also does not recognize that the solution is an explicit example of a limit cycle. More recently, Glazek and I have published two articles [35,36] showing how the limit cycle works in detail for our simple model, which is considerably easier to understand than the three-body model discussed by Bedaque *et al.* [33].

Perhaps the most fascinating possibility, for lattice gauge theorists, is a proposal that QCD could exhibit the remarkable low energy bound state structure found by Efimov, if the up and down quark masses are adjusted to produce bound states just at threshhold in both the two-body proton-neutron and di-neutron sectors [28]. The proposal is just that: a proposal. No one knows whether the Efimov behavior actually occurs for QCD. The argument that it could happen relies on a claim of universality (see [15] for a generic discussion of universality): if a renormalization group transformation is constructed for the three-body nuclear system at very low (hence non-relativistic) energies, the Hamiltonian for this transformation would iterate towards the same limit cycle as for the case of delta function potentials that has been solved explicitly. But because QCD at low energies is a strongly coupled system, there is no way of knowing in advance whether QCD, with the appropriate choice for the up and down quark masses, is in the same universality class as the delta function potential case. Otherwise it could belong to a different universality class with a totally different behavior for low energy three-body bound states.

My third topic is the understaffing of light-front QCD as an area of research complementary to lattice QCD. The heyday of light-front QCD occurred before QCD was formulated, namely in the days of the non-relativistic constituent quark model and Feynman's quark parton model. Collinear SU(6) symmetry became a basis for classifying hadrons, and the energy formula for quarks on the light front resembles the non-relativistic formula for transverse momenta although not for longitudinal momenta. But once QCD became popular, it quickly became clear that formulating this theory on the light front led to very thorny problems—the simplicity of the quark parton models could not be obtained, because of the strong coupling that meant that one would have to compute perturbation theory to very high orders, or, better, find a non-perturbative approach altogether.

But in the period from 1990 to 1994, I became part of a group of physicists at Ohio State University, including Robert Perry and Stan Glazek and various post-doctoral fellows and graduate students that made a new effort to make sense of light-front QCD in the real world of four dimensions. Our unproven "thematic presupposition" was that, for whatever reason, the QCD coupling in light-front theory never got large enough to make low orders of perturbation theory completely useless. We wanted to find a precise connection between full QCD and the much simpler picture of collinear SU(6) of quark bound states with energies determined to a considerable extent by "constituent" masses of free quarks, a contribution from an SU(6) invariant potential, and with symmetry-violating terms small enough to be treated perturbatively, yielding sum rules relating excited state energies in the presence of these perturbations. See [2] for some references. We made the assumption of a small QCD coupling at low energies out of desperation: we could not see how to get back the good features of the constituent quark model and the quark parton model unless the coupling somehow stayed at least moderately weak, even at low energy scales (below 1 GeV). We completely rethought the problems of renormalization of the light-front QCD [37], arriving at an analysis of the renormalization problem that has a number of novel and helpful features, such as a version of the renormalization group called "similarity" that is very helpful in organizing a light-front computation and an initial hypothesis about some new terms that need to be added to the Lagrangian to reflect the non-trivial vacuum of broken chiral symmetry.

After 1994 our group was weakened because both Perry and I dropped out to pursue other matters, leaving Stan and his graduate students to carry the ball, which he has done although not at the speed that a larger group could have done. But I want to indicate why the time is ripe for a few accomplished theorists to switch into light-front theory and help build a growing research effort in this area. The first question, what is our most notable accomplishment to date, is one to which I have a surprising answer. It is the simple model of a limit cycle that was proposed in the paper [34]. This paper was written as part of the early work on light-front QCD in which our group was engaged, and it illustrates that, as in the previous sections, one often cannot predict in advance what the most interesting outcomes of a research effort will be at the time it is started. I suspect that continuing research on light-front theory is likely to yield equally interesting new surprises.

But I now come back to the presupposition of our work. I can offer a simple, but bittersweet argument about the size of the coupling constant at energies below 1 GeV. It is a very simple argument in a preliminary form that is as yet unverified by any serious calculation.

The basic claim of the argument is that the running coupling constant of QCD is likely to become strong in the pure glue sector at a considerably higher value of the running cutoff than the QCD scale for quarks. Stated directly in terms of constituent masses, I believe that the best value for a constituent gluon mass is likely to be considerably higher (such as a factor of two to four) than the best values of the constituent up and down quark masses. Why? I use an analogy to the case of QED. In QED, atomic states involving 2 electrons and a helium nucleus are considerably more tightly bound than atomic states of hydrogen, simply because of the doubling of the charge in the case of helium. But in QCD, gluons belong to the octet representation, which should yield color SU(3) Clebsch-Gordon coefficients that are something like a factor of two larger than for the quark representation. This leads me to the hope that whatever the mechanism that leads to a non-zero lowest mass state in the pure glue sector (neglecting quarks) is, it would not require as large a value for the running value of $g$ as quark binding requires. This would lead to a considerably higher mass for the glueball, and hence for the constituent glue particles, than for the non-zero constituent masses of quarks.

There is a second stage to the argument, equally unproven. Namely once there is a non-zero lowest mass in the pure glue sector, this lowest mass should stop any further running of the quark-gluon coupling constant—it never gets large enough to truly bind quarks, but the quarks are bound anyway because of the linear potential between quarks if they separate very far. Hence the coupling strength between constituent quarks is relatively small, and many of the results of the constituent quark model apply because of this, from the SU(6) classification of bound states to Zweig's rule. In particular, this assumes that the gluon interactions are replaced by an effective potential between quarks that is not small (being linear) but that is invariant under SU(6); while couplings that explicitly break SU(6) are small.

This is a bittersweet result because it implies that one has to get to the very edge of perturbation theory, if not beyond it, in the pure glue sector in order to arrive at a non-zero lowest mass for the lowest mass glueball and to derive the potential of interaction between constituent quarks (including a linear behavior at long distances). But if one can find some way to do this, then one could hope to find that constituent quarks have other relatively weaker interactions that are more easily treated as perturbations relative to the bound state energies of constituent quarks in a potential. However, there would still be complications due to the presence of a considerably lower energy scale associated with the pion mass as one gets close to exact, but spontaneously broken, chiral symmetry.

There is an intriguing suggestion in the 1994 article about how spontaneously broken chiral symmetry might work in light-front QCD that is very different from normal QCD. In normal QCD, chiral symmetry is conserved only if the quark mass is zero. But in the renormalized light-front theory, this is no longer the case, at least according to our analysis, which unfortunately we can not prove: it also qualifies as a thematic presupposition. But I believe that it satisfies the Pauli criterion: our analysis in [37] might be wrong, but it is not dull.

To conclude this part of this section, I will sum up by saying that light-front QCD is not for the faint of heart, but for a few good candidates it is a chance to be a leader in a much smaller community of researchers than one faces in the major areas of high-energy physics, with, I believe, unusual promise for interesting and unexpected results.

My last topic is how to justify the expense of lattice gauge theory, indeed of high-energy physics as a whole, in a world that is confusing and changing. My answer is that one reason for governments to fund physics is the capabilities that very accomplished physicists acquire to tackle exceedingly complex problems and bring some kind of order to them. Until the end of the cold war, senior physicists with a lot of experience in military matters were able to provide very valuable and *disinterested* advice to Congress and the executive branch about technical, complex, and very expensive military matters.

But now the dominant issues facing government involve issues that economists and other social scientists study, and the government now gets more advice from social scientists than from physicists. Even military issues are hard to separate from social issues, as "winning peace" becomes as much an issue as winning wars. Moreover, while the physicists who provided military advice had considerable experience with military issues upon which to draw for their advice, there are as yet few physicists who know enough about the present state of social science to be of as much help as they might be with more knowledge and experience. I suggest that high-energy physicists might give some thought to what are the most daunting social and economic problems of today, and whether there could be any way to recruit and assist a few exceptionally capable physicists to learn considerably more about these problems.

I make this suggestion because my own recent experience with the present state of historical research has opened my eyes to the history of physicists making unique contributions to dominant issues facing society at large, from Benjamin Franklin's role in the founding of the US as a nation to the less well-known but very interesting history of physicists helping to advance such key technologies as electric power about a century ago [38].

## 7. CONCLUSION

In summary, I hope I have provided a sense of how my personal experiences contributed to developments in the 1960's and 1970's by the community of researchers in high-energy physics and in critical phenomena that existed at that time. I hope I have conveyed a sense of how limited was my accomplishment of that time, at least in lattice gauge theory, compared to what is now known. I have raised the issues of whether my accomplishments were "discoveries waiting to happen" that would have eventually been accomplished by someone else if my research had gone in a different direction. I have discussed shortcomings in my own work, and the extent to which subsequent research has overcome these shortcomings. I have offered a proposal to establish a "Pauli list" of seemingly questionable proposals that deserve more respect than they might otherwise receive, based on experiences from that of early Copernicans to those of advocates of quarks in the 1960's. But I hope I have also conveyed a sense of longer-term progress in the physics of gauge theory that could underlie a healthy research enterprise for a considerable time into the future, with unexpected benefits to society arising partly from leading elementary particle physicists who apply their unusual experience in the solving of complex problems to problems of key importance for society as a whole.

## ACKNOWLEDGEMENTS

I am grateful to the organizers of this Conference for the invitation to speak, and to the local organizing committee, especially Andreas Kronfeld and Paul Mackenzie, for all their help with the talk itself and with its written version. I thank John Laiho, Masataka Okamoto, and Ruth Van de Water for a critical reading of the manuscript. I am grateful also to George Smith, the Acting Director of the Dibner Institute, and his staff for the invitation to become a visiting Fellow of the Institute, and to George Smith, Sam Schweber, John Burnham, and Jeff Horn for extensive discussions about the history of science and its possible relevance to problems of here and now. George Smith provided helpful comments on the manuscript. Many other historians, too numerous to mention, have been helpful also. I am grateful to Constance Barsky, my collaborator on the historical research, who also was awarded a fellowship. I am grateful to Robert Perry, Stan Glazek, and others for the collaboration that led to some of the results reported here on limit cycles and on light-front QCD. I am happy to acknowledge the unfailing financial support of The Ohio State University, through a Chair fund, for research conducted prior to and after the Fellowship. In addition, during the period from 1961 to 1987, much of my research was generously supported by the National Science Foundation.